\documentclass[amsmath,amssymb,twocolumn,pre,floatfix]{revtex4}

\usepackage{siunitx}

\usepackage{color}
\usepackage{graphicx}
\usepackage{dcolumn}% Align table columns on decimal point
\usepackage{times}
\usepackage{soul}
\usepackage{hyperref}

\usepackage{afterpage}

\begin{document}

\title{Fiber-based shape-morphing architectures}
\author{Andrei Zakharov,\textit{$^{a,b}$} Len M. Pismen,\textit{$^{a}$} and Leonid Ionov\textit{$^{c}$}}
\affiliation{\textit{$^{a}$~Technion - Israel Institute of Technology, Haifa 32000, Israel}}
\affiliation{\textit{$^{b}$~University of California, Merced, 5200 N Lake Rd., Merced, California 95343, USA }}
\affiliation{\textit{$^{c}$~Faculty of Engineering Science, Ludwig Thoma Str. 36A, 95447 Bayreuth, Germany; Bavarian Polymer Institute, Universit{\"a}tsstr. 30, 95440 Bayreuth, Germany}}
\email{Leonid.Ionov@uni-bayreuth.de} 

\begin{abstract}
	We describe a combined experimental and theoretical investigation of shape-morphing structures assembled by actuating composite (Janus) fibers, taking into account multiple relevant factors affecting shape transformations, such as strain rate, composition, and geometry of the structures. Starting with simple bending experiments, we demonstrate the ways to attain multiple out-of-plane shapes of closed rings and square frames. Through combining theory and simulation, we examine how the mechanical properties of Janus fibers affect shape transitions. This allows us to control shape changes and to attain target 3D shapes by precise tuning of the material properties and geometry of the fibers. Our results open new perspectives of design of advanced mechanical metamaterials capable to create elaborate structures through sophisticated actuation modes.
\end{abstract}
 
 \maketitle
 
\section{Introduction}
Fiber-based structures are ubiquitous and can be found at different length scales, from the cytoskeleton comprising protein filaments to deployable structures in architectural engineering. At the mesoscale level, a typical example is given by spider webs that are hierarchically ordered and mainly composed of structural radial threads and sticky spiral threads (orb webs). Their lightness places these webs in the same class of structures as cable and membrane systems. Because of their unique structure and properties, spider webs became a source of inspiration for a number of industrial applications\cite{Zarek}. Even though webs do not change their shape, and their geometry is static, mechanical properties of webs may vary with humidity and temperature. Another fascinating kind of fibrous structures are tensegrity structures \cite{Pugh,Zhang}, which consist of rigid elements (rods) connected by tensioned fibers without touching each other, so that changing position and orientation of rods requires applying a load to deform (stretch) the fibers. As a result, the whole structure behaves as an elastic element if fibers are fixed at the points of contact with rods. Thus, design of morphable fibrous architectures, which are capable to reversibly change/adapt their structure/shape in response to changes of the environment, can open new perspectives for design of materials with new properties -- mechanical metamaterials.

The essential elements of webs and tensegrity structures are fibers, which form quadrilateral or triangular cells. This suggests that shape-morphing structures with tunable geometry of quadrilateral cells can be fabricated using fibers with shape-changing behavior. Although there are many materials with shape-changing properties \cite{Jeon,Stoychev,Gracias,Wu}, most of them are not suitable for fabrication of free-standing shape-changing web-like structures \cite{Stoychev}. For example, while hydrogels \cite{Ionov13} have the largest amplitude of actuation, their usage is limited to wet environments. Hydrogels are also too soft (with the typical elastic modulus in the range 102 - 105 Pascal) and too brittle to form stable fibers and webs. Shape-memory polymers \cite{Lendlein} and liquid crystalline elastomers are stiffer than hydrogels and can form long fibers which do not rupture under their own weight, and can be actuated in both dry and wet environments. On the other hand, shape-memory polymers require manual pre-deformation from a permanent to a temporary shape, while liquid crystalline elastomers \cite{Liu} actuate at relatively high temperatures \cite{Guin}, which restricts practical application of these materials. Moreover, all these shape-changing polymers are cross-linked networks because crosslinking is essential for reversibility of actuation, and their fabrication by fiber spinning and 3D printing is problematic. 

Utilization of semicrystalline polymers \cite{Ionov17} for design of shape-morphing structures offers a number of advantages: their actuation is independent of environment (dry or wet), they are able to actuate at relatively low temperatures \mbox{($< 50^{\circ}$C)}, and they undergo a reversible transition \cite{Ionov17,Behl,Qian}. These polymers are not chemically crosslinked, and reversible actuation is provided by physical entanglement between polymer chains. This makes possible their spinning from solutions and melts. However, the main limiting factor is a relatively small linear size change upon actuation, in the order of 1-5\%. Those are typical values of the cubic root of the volume change at the first-order phase transition, which depend on the degree of crystallinity of a polymer and are much smaller than the respective values for hydrogels. Thus, increasing the actuation amplitude is very important for practical application of actuating structures based on semicrystalline polymers. 

A promising way to increase the actuation amplitude is to use special geometries. The simplest way is to construct bilayer structures comprising actuating semicrystalline and passive polymers, which are similar to the bimetal structures. Such bilayers are able to bend in the direction determined by expansion or contraction of an actuating polymer \cite{Ionov17}. In other words, the bilayer geometry allows a conversion of a simple uniaxial actuation (in-plane deformation) into an out-of-plane deformation (bending). Timoshenko became the first to describe bending bilayers, mathematically, and showed that the bending curvature depends on mechanical properties of materials, their thickness and strain mismatch \cite{Timoshenko}. As a result, a few percent of linear elongation or contraction can be converted into a considerable angular deflection. However, shape transformation of bilayers is limited to simple bending, whereas combining shape-changing and passive fibers allows one to circumvent this issue and generate complex, predictable shape transformations \cite{PRM,PRE18}. In this paper, we investigate experimentally and theoretically the variety of shape transitions in structures actuated by composite Janus fibers (named by the two-faced Roman god). Such structures have a fast response time, well-controllable mechanics, and mimic an elementary quadrilateral cell of the web architecture.

The rest of this paper is organized as follows. In Sect. II we describe the experimental technique. Sect. III gives the description of bending regimes and the mechanics of elastic deformations, first for the simplest case of separate multilayered fibers and then for rings and frames. We also examine the change in reshaping of frames caused by the reorientation of the Janus fibers, demonstrating the possibility of attaining multiple shapes of the same frames.

\section{Sample fabrication}

The general procedure for fabrication of bilayer fibers involves 3D printing with layer-by-layer fusion. The fibers are printed using a commercial desktop fused filament fabrication 3D printer with two separate extruders having standard nozzles of inner diameter 0.3 mm. Print paths are generated by executing G-codes to control the three-dimensional motion of the nozzles. A glass slide covered by polyimide tape and heated to 60$^{\circ}$C is used as a support.

As the passive layer, we use polylactide (PLA, Janbex Co., Germany), an amorphous glassy polymer with a high temperature of glass transition ($T_g = 65^{\circ}$C) and a high melting point  $T_m = 175^{\circ}$C. PLA is extruded at a constant feed rate from a nozzle heated to 220$^{\circ}$C and solidifies before printing the next layer. The actuating material is polycaprolactone (PCL, Sigma-Aldrich Corp., USA) with the molecular weight 45 and 80 kDa and a low melting point $T_m = 60^{\circ}$C. The filament is custom fabricated from polymer beads using a desktop extruder at 100$^{\circ}$C, and is extruded from the printing nozzle heated to 200$^{\circ}$C.

Depending on the elevation of the nozzle, the thickness of each printed layer varies from 0.1 to 0.4 mm. The fiber width depends on both the extrusion rate and the nozzle speed. These parameters are chosen in a way allowing us to print fibers with a cross-section close to a square. The printed structures are cooled to the room temperature after printing, and easily detach from the glass support plate when actuated.

The mechanical properties of the above polymers are as follows: the Young moduli $E_{PLA}=\SI{1.2}{\giga\pascal} $ and $E_{PCL}=\SI{0.3}{\giga\pascal}$ (measured at 20$^{\circ}$C using Anton Paar MCR 702 Multidrive Rheometer, Anton Paar GmbH, Austria), the densities $\rho_{PLA}=\SI{1.25}{\gram\per\cubic\centi\metre}$ and $\rho_{PCL}=\SI{1.15}{\gram\per\cubic\centi\metre}$.

\section{Results and discussion}

\subsection{Properties of Janus fibers}

\begin{figure}[t]
	\centering
	\includegraphics[width=0.48\textwidth]{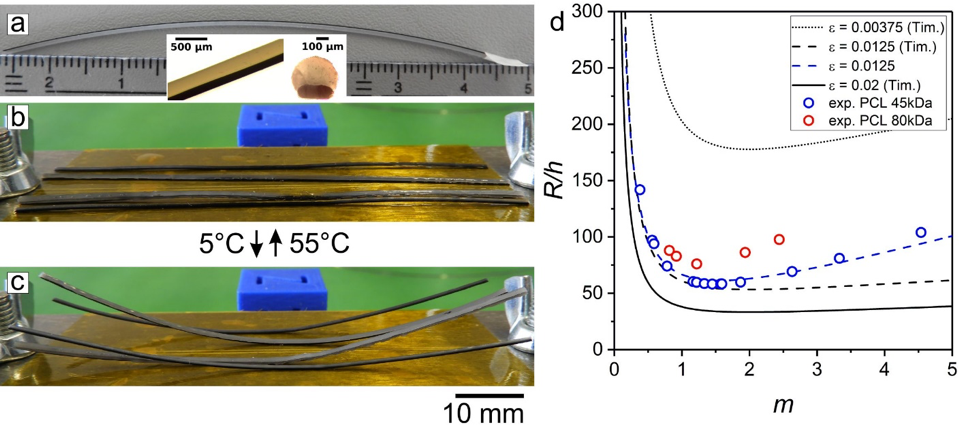}
	\caption{Shape-changing properties of individual Janus fibers. (a) A segment of a Janus fiber at the room temperature. (b,c) Experimental results of actuation of Janus fibers of different thickness caused by contacts with a hot (b) and cold (c) plate. (d) Dependence of the dimensionless actuation amplitude $R/h$ on the ratio $m = h_{PCL}/h_{PLA}$ between the two polymers constituting the fiber. The black dotted, dashed, and solid lines correspond to the theoretical predictions of $R/h$ at the constant $h_p = 0.3, n=0.25, w_p = 0.625, w_a = 0.8$, and different values of the extension coefficient $\varepsilon$ according to the Timoshenko equation (\ref{Te}). The blue dashed line shows the theoretical prediction of $R/h$ at $h_p = 0.3, n=0.25 w_p = 0.625, w_a = 0.8, \varepsilon = 0.0125$ according to our model equation (6). The red and blue open circles show the experimental data for open-ended Janus fibers made of PCL with different molecular weight at $n\approx0.25$ and $w_p\approx0.625$ mm, $w_a \approx0.8 $ mm.}
	\label{fig:JF}
\end{figure}

First, we investigated actuation of individual unconstrained PCL/PLA bilayer fibers (see Fig.~\ref{fig:JF}a) with the cross-section shown in the inset . At the temperature close to the melting point of PCL, the fibers are relaxed and straight (Fig.~\ref{fig:JF}b). When temperature decreases, PCL contracts, and the fibers bend towards the PCL side. Consequently, the fibers develop a curvature with the center of curvature on the PCL side (Fig.~\ref{fig:JF}c). The experimental results show that an increase of the ratio between the thicknesses of PCL ($h_{PCL}$) and PLA ($h_{PLA}$) layers causes, first, an increase, and then, a decrease of the curvature. We fabricated fibers differing by their total thickness $h = h_{PCL}+h_{PLA}$ and the ratio between the thicknesses of its components $m = h_{PCL}/h_{PLA}$, in order to find optimal parameters leading to the maximal actuation amplitude, and investigated the dependence of the bending curvature on the fiber geometry. According to the classical result of Timoshenko for bimetallic strips \cite{Timoshenko}, the curvature radius of a bilayer R is proportional to the thickness h, thus, the extension coefficient can be related to the actuation amplitude measured by the dimensionless parameter $R/h\gg1$. We measured the curvature of fibers observed at the room temperature after 10 cycles of heating/cooling (open circles in Fig.~\ref{fig:JF}d). Since the layers are made of different polymers, they have different flexural rigidities accounted for by the ratio , where , are the elastic moduli of PCL and PLA, respectively. We observed the minimal curvature radius ($R/h\approx 57$) at the thickness ratio $m \approx1.5$ for a fiber with PCL of the molecular weight $Mn \approx45$ kDa, The fibers made of PCL with a higher molecular weight ($Mn \approx 80 $ kDa) developed a lower actuation amplitude with a larger ratio, $R/h \approx76$, and the maximum curvature was observed at the thickness ratio $m\approx1.1$.

To extend our understanding of shape-changing regimes, we developed a theoretical model for the deformation of bi-polymer strips. The linear coefficients of thermal expansion for PCL and PLA are constant, $16 \cdot 10^{-5}/^\circ$C  and $8.5 \cdot 10^{-5}/^\circ$C \cite{Mark,Pilla} respectively, so that the difference in these values between the two polymers corresponds to the extension coefficient $\varepsilon= 0.00375$ when it is cooled (or heated) in the interval from 5$^\circ$C to 55$^\circ$C. We explored the possibilities of applying three different values of the parameter $\varepsilon$ to estimate the deflection using the classical Timoshenko equation, which assumes equilibrium conditions, i.e., deformations attaining their steady state. The Timoshenko equation can be written in form:

\begin{equation}
\frac{R}{h}= \frac{3(1+m)^2+(1+m n)(m^2+\frac{1}{mn})}{6(\varepsilon_{PLA}-\varepsilon_{PCL})(1+m)^2}.
\label{Te}
\end{equation}

For the fibers shown in Fig.~\ref{fig:JF}a with $h \approx0.6$ mm , $n\approx0.25, m \approx 1$ and $\varepsilon = 0.00375,$ the Timoshenko theory predicts $R \approx120$ mm ($R/h \approx200$) with the minimum of $R/h \approx178$ reached at $m = 2$ (black dotted line in Fig.~\ref{fig:JF}d), which is much larger than that observed experimentally ($R/h \approx57,$ open circles in Fig.~\ref{fig:JF}d). The experiments demonstrate a much stronger bending than that expected from the Timoshenko equation because the actual value of $\varepsilon$ is apparently much larger than that calculated solely on the basis of the thermal expansion coefficients, and is related to a phase transition. The approximate applicable value of $\varepsilon$ can be found by fitting the experimental and theoretical data for particular values of $h, n$ and $m$. The best fit is attained, as shown in Fig.~\ref{fig:JF}d, at $\varepsilon\approx0.0125$. This value will be used below for theoretical estimates of the actuation. Previously, we have determined that the expansion coefficient of PCL upon its melting is $\varepsilon\approx0.02$ \cite{Ionov17}, which is higher than that obtained by fitting the experimental results. 

The Timoshenko equation (\ref{Te}) considers a bilayer structure where both layers have the same width. The polymer layers in the fibers fabricated in this work have different widths, which also has to be taken into account. Generally, the energy stored in a deformed thin bilayer filament comprises stretching, bending, twisting, and gravitational contributions. The stretching rigidity is proportional to the cross-sectional area, while the bending and twisting rigidities are proportional to the cross-sectional area squared; thus, a thin filament can be assumed incompressible. For a typical fiber in our experiments with the thickness $h = 0.6$ mm, the average Young modulus $E=\SI{0.75}{\giga\pascal} $, the density $\rho=\SI{1100}{\kg\per\cubic\metre}$, so that the gravitational bending length is $l_{gb} \approx100$ mm, which exceeds the fiber length $L = 80$ mm. It is similarly to what was measured experimentally for a fiber with $h \approx0.6$ mm , $n \approx0.25, w_p \approx0.625$ mm, $w_a \approx0.8$mm, which starts to bend under its own weight when $L>130$ mm at the room temperature. Consequently, the gravitational contribution can be neglected. Under these assumptions, the equilibrium shape of a filament is therefore determined by minimizing the sum of bending and twisting energies:

\begin{equation}
\mathcal{F} =  \frac12 \int \left( I + J \right) ds, \label{Fe1}
\end{equation}
where $I, \, J$ are the bending and twist momenta, respectively. 

Since we are specifically interested in Janus fibers combining passive and actuated components, the bending momentum, determined by the total elastic energy per unit cross-sectional area of a filament, is computed by adding the contributions of the actuated ($I_a$) and passive ($I_p$) layers of different thickness, $h_a$ and $h_p$, and width, $w_a$ and $w_p$, respectively. Upon actuation, the actuated component contracts locally by the factor , causing the filament to develop internal curvature $\kappa=1/R$. The expressions for the bending momenta of the two layers are

\begin{align}
I_p &=E_p\int_{-w_p/2}^{w_p/2} 
\int_{0}^{h/(1+m)} (\kappa y)^2 \,d y \,d x = \frac{E_p w_p \kappa^2 h^3}{3(1+m)^3}, \label{Ioutz}
\\
I_a &=E_a\int_{-w_a/2}^{w_a/2} 
\int_{h/(1+m)}^{h} (\kappa y-\epsilon)^2 \,d y \,d x = \notag \\
&= \frac{E_a w_a((\epsilon-\kappa h/(1+m))^3+(\kappa h - \epsilon)^3)}{3\kappa} , \label{Iinz} 
\\
I &= I_a+I_p= \notag \\
&= \frac{E_p w_p \kappa^2 h^3}{3(1+m)^3} +\frac{E_a w_a((\epsilon -\kappa h/(1+m))^3+(\kappa h-\epsilon)^3)}{3\kappa},
\label{Ipi}
\end{align}

where again $m=h_a/h_p$ and $h=h_a+h_p$. The equilibrium curvature $\widehat{\kappa}$ is defined by the condition $dI/d\kappa=0$, leading to

\begin{equation}
\widehat{\kappa}= \frac{3\epsilon  m (m^2+3m+2) }{2h (E_p w_p/(E_a w_a) +m(m^2+3m+3))}.
\label{khat}
\end{equation}

The blue dashed line in Fig.~\ref{fig:JF}d shows the dependence of the optimal curvature radius $1/\widehat{\kappa}$ on the ratio $m$ at fixed parameters $E_a, E_p, h_p$ taken from experiment, and $\varepsilon=0.0125$ chosen for modeling as the best agreement between Eq.~(\ref{khat}) and experimental results. It is presented alongside the experimental data for the fibers with the corresponding thickness. 

\subsection{Comments on the model and its parameters}

Commonly, the actuation of bilayers is modeled using Eq.~(\ref{Te}), but Eq.~(\ref{khat}) is more precise, as it takes into account differences in the layer widths. The apparent reason why the extension coefficient obtained by fitting experimental results is lower than that measured previously \cite{Ionov17} lies in the viscoelastic behavior of non-crosslinked PCL. Volume expansion coefficients are usually measured well below the melting point. Since melting of a polymer takes place in a wide temperature range (normally 20$^{\circ}$C), the volume expansion strongly increases at temperatures close to melting. On the other hand, the expansion coefficient obtained by fitting experimental results is lower than that measured previously \cite{Ionov17}. Viscoelastic behavior of non-crosslinked PCL is the most probable reason for a lower actuation amplitude. In fact, shrinking PCL exerts force on the PLA layer during actuation leading to bending the bilayer. On the other hand, the bent PLA side exerts a force on PCL that leads to its irreversible deformation (viscous flow).

The actuation of bilayers depends on the molecular weight of PCL. The Timoshenko equation (\ref{Te}) was derived for purely elastic deformations, which are independent of the time scale of deformation. The deformation of polymer melts and solid polymers is not purely elastic but has a viscoelastic character. The viscous component of deformation retards all actuation processes but does not affect the equilibrium shape. Thus, a lower actuation amplitude of bilayers with PCL with high molecular weight (80 kDa) can be explained by its higher viscosity. High viscosity reduces the actuation speed, so that the bilayer is unable to reach its equilibrium shape before it cools down to the room temperature when it crystallizes. Using polymers with a very low molecular weight will not result in actuation because of the absence of entanglements ($M_e \approx$ 30kDa). From this point of view, using PCL with the molecular weight 45 kDa is most favorable: this polymer is able to demonstrate the elastic behavior, and its viscosity is much lower than that of PCL with the molecular weight 80 kDa. In fact, viscosity scales linearly with the molecular weight until its critical value $M_c$, which is roughly double of $M_c$(PCL) $\approx$ 60kDa. Viscosity scales exponentially with the molecular weight above $M_c$. All this means that both PCL 45 kDa and PCL 80 kDa have entanglements allowing their reversible actuation. On the other hand, PCL 45 kDa has a lower molecular weight than $M_c$ (with viscosity scaling linearly) and PCL 80 kDa has a higher molecular weight than $M_c$ so that viscosity scales exponentially. It is impossible to develop an advanced Timoshenko-based model which includes the viscous component, because it must include an empirical coefficients responsible for kinetics of deformation and the change of properties of materials in time. These coefficients should account for kinetics of crystallization and the rate of heat exchange. 

The most important observation, which can be derived from the comparison of experimental and modelling results, is an estimate of the degree of proximity of observed deformed shapes to the equilibrium state. We found that the typical actuation times for a fiber reshaping are comparable to the actuation times of fast thermally responsive polymer-based actuators (2-3 s) \cite{Hines}. Both the fiber thickness and the molecular weight of PCL have a direct bearing on the actuation times. Thick fibers actuate slower because they demand more time to be uniformly heated/cooled. The relaxation time scales in polymers with the molecular weight as $M^{3.4}$, which results in a slower reorganization in PCL with a higher molecular weight that cannot completely contract reaching the equilibrium volume before cooling down to the room temperature. Similar processes are also observed in PCL with a lower molecular weight (45 kDa), but, due to a higher molecular mobility, PCL is able to undergo stronger deformations before cooling down to the room temperature.

\subsection{Janus rings}

\begin{figure}[t]
	\centering
	\includegraphics[width=0.48\textwidth]{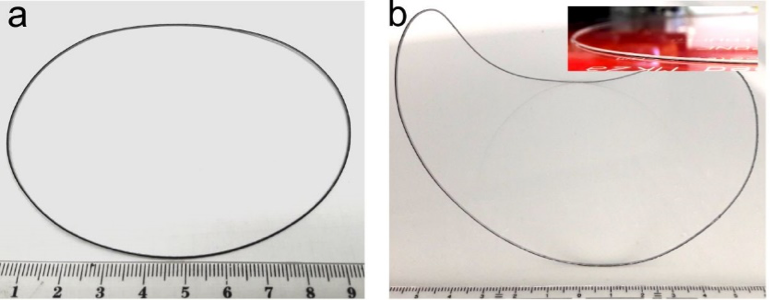}
	\caption{Shape transformation of a bilayer ring: (a) A planar Janus ring of a small radius ($R_0 = 50$ mm). (b) The equilibrium configuration of a ring with $R_0 = 80$ mm at the room temperature and on a hot plate (inset).}
	\label{fig:JR}
\end{figure}

Next, we investigated deformation of bilayer fibers with a constrained geometry -- a ring, which mimics radial fibers of webs. In contrast to  bilayer beams with open ends discussed above, the ring is closed, which substantially constrains its actuation. Closed bilayer fibers \cite{PRE18} were printed as circles of the radius $R_0$ within the range 50 - 80 mm, and the elevation of the nozzle for each layer 0.125 mm and 0.225 mm, respectively, yielding the total fiber thickness $h = 0.35$ mm and the thicknesses ratio of $m\approx1.8$ (Fig.~\ref{fig:JR}a). It corresponds to the intrinsic curvature radius $R \approx24 $mm at $\varepsilon=0.0125$ upon actuation in an unconstrained configuration (Fig.~\ref{fig:JF}d).  In the reference state at a high temperature, the curvature is oriented normally to the surface (Fig.~\ref{fig:JR}b), because the fiber is printed layer-by-layer. We found that a ring of a small radius ($R_0 < 80$ mm) remains planar when cooled (Fig.~\ref{fig:JR}a), in spite of contracting of the PCL layer. Since the ratio $R_0/R < 3.33$ is small, it leads only to a uniform rotation of the dividing plane between the two layers about the fiber axis, so that the contracting PCL layer becomes located inside the ring. When $R_0 = 80$ mm, the incompatibility between the internal curvature and the ring radius causes the ring to deviate from the original plane in order to reduce the bending energy. This causes the fiber to twist (Fig.~\ref{fig:JR}b), and the orientation of the dividing plane varies along its length. Earlier theoretical studies showed \cite{PRM,PRE18} that the shape of a Janus ring is governed by the ratio $R_0/R$ dependent on the extension coefficient, and can exist in multiple states, following discontinuous transitions between shapes with a different number of "petals". The linear stability analysis of an untwisted ring predicts the following numerical values $R_0/R = 3.2, 6.6$ and $11.3$ for the transitions to shapes with 2, 3, 4 petals, respectively. In our experiment the planar shape becomes unstable at the ratio $R_0/R \approx3.33$ producing two petals, which is in a good agreement with the theoretical results. The three-dimensional reshaping between different stable states depends on the expansion coefficient $\varepsilon$, which affects $R$, as well as on the initial radius of the ring $R_0$. For example, further increasing the expansion coefficient or the ring radius ($R_0/R > 3.2$) is expected to result, first, in a rising amplitude of petals, and then in increasing their number at $R_0/R > 6.6$. Decreasing the fiber diameter should result in a proportional reduction of $R$ and lead to transitions to a larger number of petals at a constant length. Individual rings can be potentially connected to each other to form structures with a complex shape-morphing behavior. 

\subsection{Woven frames}

Next, we investigated shape transformation of structures consisting of several bending bilayer fibers and connecting passive (non-actuated) fibers. As an illustrative example of shape-changing architecture, we tested frames combining Janus fibers and monolayer passive fibers. Such frames can be considered as elementary lattice units in webs. The passive fibers are made of PLA, they are homogeneous, and have a circular cross-section of the radius $r_p = 0.3$ mm. The Janus fibers have the overall thickness 0.7 mm and are printed by fusion of one PLA layer with the thickness $h_p = 0.27$5 mm and width $w_p = 0.8$ mm and two PCL layers of the total thickness 0.425 mm and the width 0.625 mm. In this geometry, the Janus fibers have the intrinsic curvature radius about 55 mm attainable upon actuation with no constraints (as in Fig.~\ref{fig:JF}d). In the reference hot state, all fibers are straight and have the same length $L = 78$ mm. We worked with macroscopic fibers because of the simplicity of experiments. All obtained results will also be valid for smaller fibers with the same ratio between the radius of curvature and length. According to the Timoshenko equation (\ref{Te}), a smaller radius of curvature can be achieved when thinner fibers are used. Thus, thin and short fibers will produce the same geometries but reduced in size. 

\begin{figure}[t]
	\centering
	\includegraphics[width=0.48\textwidth]{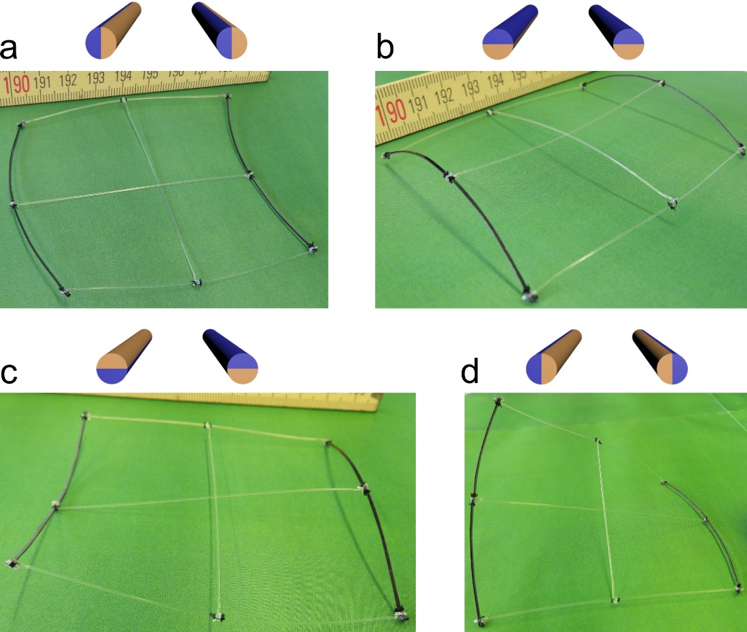}
	\caption{A frame actuated by two Janus fibers (black) oriented (a,b) in the same and (c,d) in opposite direction. Cartoons on the top of each image represent schematically the orientation of the Janus fibers.}
	\label{fig:WF}
\end{figure}

\begin{figure}[t]
	\centering
	\includegraphics[width=0.48\textwidth]{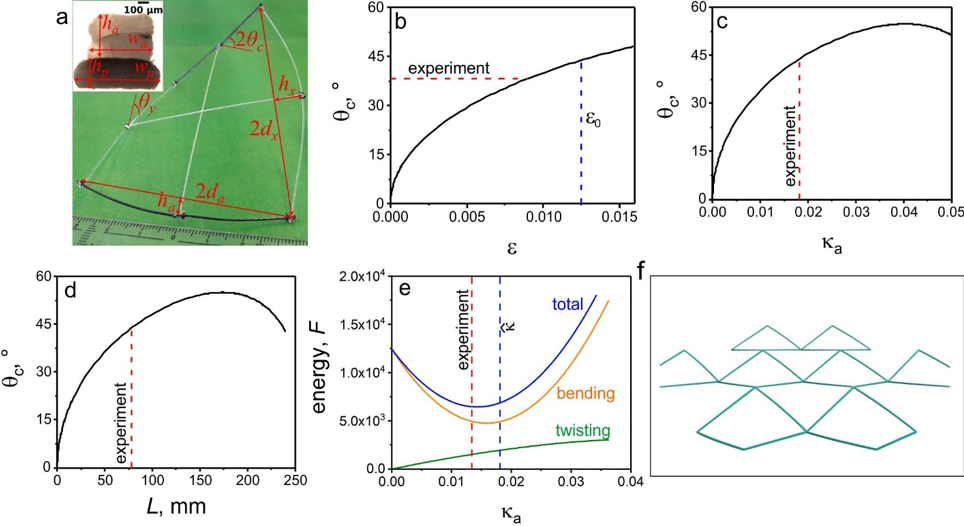}
	\caption{(a) A sketch of the geometrical parameters of a frame and a cross-section of a Janus fiber (inset). (b-d) The twist angle $\theta_c$ as a function of the extension coefficient $\varepsilon$, the curvature $\kappa_a$, and the fiber length $L$. The vertical dashed line in (b) corresponds to $\varepsilon_0 =0.0125$ and $\theta_c\approx43^{\circ}$  for an open-end fiber, while the horizontal dashed line shows the experimentally observed angle $\theta_c\approx37^{\circ}$. (e) Dependence of the dimensionless bending, twisting and overall energy of the frame on the curvature of the Janus fibers. The vertical dashed lines show the optimal and experimental values of the curvature $\kappa_a$. Parameters for (b,c,d,e): $L=78, E_a = 0.3\cdot 10^{9}, E_p = 1.2\cdot 10^{9}, m = 1.545, h = 0.7, w_a = 0.625, w_a = 0.8, r_p = 0.3$. (f) 3D view of a possible multiframe structure with negative Poisson ratio behavior.}
	\label{fig:WFT}
\end{figure}

We assembled first a frame containing two Janus and four passive fibers connected by loop joints at their ends (Fig.~\ref{fig:WF}), which allows the Janus fibers to freely rotate around their axes without a twist, and to deform into a stable state upon actuation. The passive fibers crossing each other in the center are allowed to detach. Since the structure is sparse and the Janus fibers are allowed to rotate, there are multiple equilibrium configurations but there is no state with the Janus fibers reaching the optimal curvature and the passive filaments remaining undeformed. For instance, when the Janus fibers develop curvature in the same direction, either the framing passive or the parallel central fiber must bend, because the distance between the ends of Janus fibers shortens. Depending on the initial state, the frame remains planar (Fig.~\ref{fig:WF}a) or undergoes a 3D deformation (Fig.~\ref{fig:WF}b-d).

Configurations with the opposite orientation of the Janus fibers always reshape out-of-plane, and their bending induces twist of passive filaments. These configurations are metastable, and are attainable only if the Janus fibers are constrained to retain their orientation upon actuation.  For this purpose, we firmly connect them to the central perpendicular fiber. When the Janus fibers bend normally to the frame, the central passive filaments do not deform, while the framing ones become both bent and twisted (Fig.~\ref{fig:WF}c). All fibers are deformed, and the frame has a maximum deviation from the reference state if the Janus fibers are set to bend parallel to the structure's plane (Fig.~\ref{fig:WF}d). In this case, the distance between the ends of the Janus fibers shortens, and, consequently, the frame twists about the central fibers to reduce the curvature of the framing passive filaments. We will consider the shape-changing behavior in detail analytically below.

To gain further insight into the origin of out-of-plane actuation detected in the experiments, we investigated the interplay between bending and twisting energies, which are difficult to distinguish experimentally. For this aim, we applied the analytical theory \cite{PRM} to the results presented in Fig.~\ref{fig:WF}d for a frame with a symmetric actuation of Janus filaments developing curvature in the frame plane, and demonstrate the transition to a twisted shape.

The energy of the structure is dependent on the curvatures $\kappa_i$  of the driven, parallel passive, and perpendicular framing and central filaments (marked by the indices $i (a, y, x, c)$, respectively) and the twist angles $\theta_x, \theta_c, \theta_y$ (Fig.~\ref{fig:WFT}a). The Janus fibers are not clamped at ends, so that no torque is applied, and the twist angle $\theta_a=0$. Since the curvature radii of the Janus fibers are oriented inside the frame, both central filaments remain straight $\kappa_c = \kappa_y = 0$.  Thus, the general expression for the energy per unit volume can be written as:
\begin{align}
\mathcal{F}=& 2(\mathcal{F}_a^{b}+\mathcal{F}_x^{b})+2\mathcal{F}_x^{t}+\mathcal{F}_c^{t}+\mathcal{F}_y^{t},
\label{engen}
\end{align}
where $\mathcal{F}_x^{t}=\mathcal{F}_c^{t}$ due to the symmetry. The bending and twist energies for homogeneous passive fibers with circular cross-section of radius $r$ have a standard form $\mathcal{F}_x^{b}=E_p \pi r^4 \kappa_x^2/8$ and $\mathcal{F}_i^{t}=E_p \pi r^4 (2\theta_i/L)^2/12$. 

To detect the dependence of the curvature and twist of the passive filaments on $\kappa_a$ and $\epsilon$, it is convenient to define arc angles $\phi_i=L_i\kappa_i$ with the lengths of all passive filaments equal to $L$ and $L_a=L(1-\epsilon/2)$. We will also need the chord half-lengths (half-distances between the ends of bent fibers) $d_i=\sin (\phi_i/2)/\kappa_i$ and the heights of the segments (sagitta) $h_i =(1-\cos (\phi_i/2))/\kappa_i$ . Rotating the Janus filaments off-plane around their midpoint in opposite directions by some angle $\theta_c$ reduces the projection of the base of the filament on the original plane to $\widehat{d}_a=d_a\cos \theta_c$ as a result the length of passive framing filaments conserves but curvature $\kappa_x$  increases. On the condition that the central filaments are straight, the curvature $\kappa_x$, as well as the twist angles $\theta_c,\theta_y$ can be found by solving the relations:

\begin{gather}
h_x =|d_y-\widehat{d}_a|,  \qquad  d_x^2=(L/2-h_a)^2+(d_a \sin \theta_c)^2  \notag \\
\sin \theta_y = d_a \sin \theta_c/d_x.
\label{hxdx}
\end{gather} 

In Fig.~\ref{fig:WFT}b we present the dependence of the twist angle $\theta_c$ on $\epsilon$ at $\kappa_a = \widehat{\kappa}$. The expected value of the angle at $\epsilon_0=0.0125$ (the vertical dashed line), which corresponds to the actuation amplitude for an unconstrained Janus fiber, is larger than the angle detected experimentally (the horizontal dashed line), $\theta_c\approx43^\circ$ (expected) and $\theta_c\approx37^\circ$ (measured), respectively. The difference is caused by the twist energy that rises with $\epsilon$ and $\kappa_a$. However, if the fibers are connected by movable joints, there is no applied torque, and the frame is not twisted. The influence of the twist energy and a trade-off between the elastic energies to attain equilibrium shape with a minimal overall energy becomes more pronounced when they are plotted against $\kappa_a$  at the fixed $\epsilon_0=0.0125$ (Fig.~\ref{fig:WFT}c). The length of the fibers also affects the twist angle: an increase of length results in an increase of the twist angle, until a certain value is achieved (Fig.~\ref{fig:WFT}d). A further length increase results in a decrease of the twist angle, since the active fibers reach the shape of a semicircle and the distance between their ends shrinks. One can see that the condition $\kappa_a = \widehat{\kappa}$ is not optimal, and the total energy $\mathcal{F}$ (Fig.~\ref{fig:WFT}e) has a minimum at a lower $\kappa_a=0.0143$, in a good agreement with the experiment ($\kappa_a=0.0138$). The bending energy has a minimum at $\kappa_a$ between the observed curvature and $\widehat{\kappa}$, because the Janus filaments cannot attain the optimal curvature without bending passive framing fibers. The twist energy gradually grows with $\kappa_a$ and prevents the Janus fibers reaching their optimal shape. The shape-changing frames can be potentially connected together to form a web-like architecture, which would be able to change its morphology (Fig.~\ref{fig:WFT}e). This structure is expected to be stable because there is only one degree of freedom -- twisting about the central fiber.

A schematic illustration of possible structure comprising multiple actuating frames is shown in Fig.~\ref{fig:WFT}f. Each frame has a structure depicted in Fig.~\ref{fig:WFT}a and is connected by flexible joints to other frames at corners forming a lattice arranged in staggered rows. This structure reshapes from the planar state to a 3D structure that shrinks in the $x$ direction when its elements twist. An applied external force pulling in the $y$ direction will lead to a decreasing twist and an extension in the $x$ direction. This corresponds to an effective negative Poisson ratio for the multiframe structure.

%\begin{figure*}
% \centering
% \includegraphics[height=3cm]{example2}
% \caption{A two-column figure.}
% \label{fgr:example2col}
%\end{figure*}

\section{Conclusions}

Although semicrystalline polymers have a limited actuation rate, a combination of actuating layered fibers and passive components results in strong deformations. Our integrated experimental and theoretical study of shape transformations of ring and frame structures reveals their complex shape morphing behavior. We examine deformations taking into account multiple relevant factors, such as the extension ratio, the thickness and elastic moduli of layered Janus fibers, and their arrangement. This demonstrated a good agreement between the experiment with 3D-printed fibers and theoretically calculated shapes allowing one to control the shape changes and to attain target 3D shapes by precise tuning of the material properties and fiber geometry. Our study provides an example of theoretical prediction of properties of objects formed by real materials: all the parameters used for modeling are experimental values measured for the relevant materials. Straightforward computation experiments allow us to predict twisting angles and lowest-energy configurations of frame structures. Our theory can be extended to more complex structures \cite{PRM} and more sophisticated ways of actuation. The shape transformation was studied here for centimeter-scale macroscopic structures but the results are scalable and remain valid for fibers with proportionally reduced geometric parameters. The investigation of shape transformation of relatively simple shape-changing frames is the first step toward the development of elaborate shape-morphing webs, tensegrity structures and textiles, expected to undergo advanced shape morphing scenarios.

\section*{Acknowledgements}
LI is thankful to DFG (grants IO 68/10-1 and 68/11-1) for financial support. AZ and LMP acknowledge the support by the Israel Science Foundation (grant 669/14). The authors thank Mr. Andreas Posada for measuring the mechanical properties of polymers.

%%%END OF MAIN TEXT%%%



\end{document}